# Duality of momentum-energy and space-time on an almost complex manifold

You-gang Feng

Department of Basic Sciences, College of Science, Guizhou University, Cai-jia Guan, Guiyang, 550003 China

Abstract: We proved that under quantum mechanics a momentum-energy and a space-time are dual vector spaces on an almost complex manifold in position representation, and the minimal uncertainty relations are equivalent to the inner-product relations of their bases. In a microscopic sense, there exist locally a momentum-energy conservation and a space-time conservation. The minimal uncertainty relations refer to a local equilibrium state for a stable system, and the relations will be invariable in the special relativity. A supposition about something having dark property is proposed, which relates to a breakdown of time symmetry.

1. Introduction

   The uncertainty relations are regarded as a base of quantum mechanics. Understanding this law underwent many changes for people. At fist, it was considered as a measuring theory that there was a restricted relation between a position and a momentum for a particle while both of them were measured. Later on, people further realized that the restricted relation always exists even if there is no measurement. A modern point of view on the law is that it shows a wave-particle duality of a microscopic matter. The importance of the uncertainty relations is rising with the development of science and technology. In the quantum communication theory, which has been rapid in recent years, an uncertainty relation between a particle's position and a total momentum of a system should be considered while two particles' entangled states are used for communication[1]. Furthermore, an uncertainty relation is discussed in a higher-dimensional space[2]. Even though in the string theory or the quantum-gravitation theory, of which both need to be developed and perfect further, the relations are playing an important role[3–11]. The time and positions and the energy and momenta in the relations consist of two 4-dimensional spaces, respectively, that is, the space-time and the momentum-energy, such that we believe the relations are certainly concerned with some restrictions between these two 4-dimensional spaces.

In this paper we try to investigate a spatial meaning for the uncertainty relations, mainly for the minimal uncertainty ones, from the point of view of the construction of these two spaces.

In quantum mechanics an observable dynamical variable simply associates to an operator, and the variable should be a real number[12]. The position-momentum uncertainty relations can be verified by means of the wave equation because of a non-commutability of a momentum operator with a position operator. The energy-time uncertainty relation, however, can't be simply derived from the operators, since there not be a concept of time operator in the quantum mechanics at all. A typical indirect verification for the relation is that a wave packet, a wave-group velocity and a particle's energy are considered for a single-color plane wave of a free particle, and it can then be proved by means of the position-momentum uncertainty relation[13]. The reference [12] points out that a defect of this method is that the wave function used in the verification is a very exclusive-particle one such that a generality is, more or less, lacking in the derivation. It is much more general to think of a wave function depending on time. Another method to use operators in the proof is employed, in which an operator for an observable variable A is introduced provided that the energy operator H, Hamiltonian, does not directly depend on time, the energy-time uncertainty relation is then verified by means of the Schuwatz's inequality[12]. But the introduced operator A has no a direct meaning of time, and it seems that there may be no harmony between this method and the method applied in the verification of the position-momentum uncertainty relations. For this reason, some scholars doubt a correctness of this method[14,15], and there then are some new thinking routes to improve the proving of the energy-time uncertainty relation, in which a time operator is then introduced[16–19]. So far, however, there has not been any method to derive both of the position-momentum uncertainty relations and the energy-time one simultaneously, in a point of view of a restricted relation between these two 4-dimensional spaces. We think that an importance of the above opinion is that these two 4-dimensional spaces not only are basic spaces for matter motion, but also are essential reference frames for observing moving matter. On one hand, the particles move in the spaces; on the other hand, these spaces restrict to each other through the uncertainty relations, which is such that the law should reveal a nature of these two spaces in structure.

## 2. Theory

**2.1. Probability density and real manifold under quantum mechanics**

Let $\Psi(t,x,y,z)$ be a wave function describing a particle, a probability density $\omega(t,x,y,z)$ near a point $(x,y,z)$ at $t$ be given by

$$\omega(t,x,y,z) = \Psi(t,x,y,z)\Psi^*(t,x,y,z) = |\Psi|^2 \qquad (1)$$

where the wave functions $\Psi(t,x,y,z)$ and $\Psi^*(t,x,y,z)$ satisfy the wave equation and its conjugate one

$$i\hbar \frac{\partial \Psi}{\partial t} = -\frac{\hbar^2}{2m}\nabla^2 \Psi + U(x,y,z)\Psi \qquad (2.1)$$

$$-i\hbar \frac{\partial \Psi^*}{\partial t} = -\frac{\hbar^2}{2m}\nabla^2 \Psi^* + U(x,y,z)\Psi^* \qquad (2.2)$$

Note that the $U(x,y,z)$ represents a potential-energy operator, which is a real valued function and is abbreviated by $U$ as follows. From (2.1) and (2.2), we further have

$$\frac{\partial \Psi}{\partial t} = i\frac{\hbar}{2m}\nabla^2 \Psi - i\frac{1}{\hbar}U\Psi \qquad (3.1)$$

$$\frac{\partial \Psi^*}{\partial t} = -i\frac{\hbar}{2m}\nabla^2 \Psi^* + i\frac{1}{\hbar}U\Psi^* \qquad (3.2)$$

The variable relation of the probability density $\omega$ is

$$\frac{\partial \omega}{\partial t} = \Psi^* \frac{\partial \Psi}{\partial t} + \frac{\partial \Psi^*}{\partial t}\Psi \qquad (4)$$

By (3.1), (3.2) and (4), we then obtain

$$\frac{\partial \omega}{\partial t} + \nabla \cdot \vec{J} = 0 \qquad (5)$$

where

$$\vec{J} = i\frac{\hbar}{2m}(\Psi \nabla \Psi^* - \Psi^* \nabla \Psi) \qquad (6)$$

The (6) is of a form of continuous dynamical equation[20]. Since $|\Psi|^2$ is a real-valued function, the $\partial/\partial t$, the $\partial/\partial x$, $\partial/\partial y$ and $\partial/\partial z$ in the operator $\nabla$ can simply be regarded as bases for a tangent vector space to the Euclidean space $(t,x,y,z)$ called a coordinate representation $(x, y, z)$ with a time parameter $t$, the tangent space is flat and can be established at all points $(t, x, y, z)$. These bases have not any physical meaning, and its dual bases $(dt, dx, dy, dz)$, however, the bases of a cotangent vector space to the space $(t, x, y, z)$ is then denoted as bases of a space-time of dimensional of 4 in physics, which meaning will be kept even in the special relativity theory.

The following orthonormal conditions hold for these bases

$$\frac{\partial}{\partial t} dt = 1 \quad , \quad \frac{\partial}{\partial \alpha} d\alpha = 1 \tag{7}$$

where $\alpha = x, y, z$.

## 2.2. Duality between momentum-energy and space-time on an almost complex manifold

In a differential geometry sense, we can construct a tangent vector space by some differential operators, and its dual space is a cotangent space such as the familiar 4-dimensional space-time. In quantum mechanics an energy and a momentum all are in 1:1 correspondence to the operators given by

$$E \to i\hbar \frac{\partial}{\partial t} \quad , \quad P_\alpha \to -i\hbar \frac{\partial}{\partial \alpha} \tag{8}$$

where the constant $\hbar$ is a quantum symbol, the imaginary number i indicates that these operators, as bases, construct an almost complex space[21], which is an almost complex structure on a real vector space mentioned in the subsection **2.1.**. Such a space is an incomplete complex space containing only an imaginary part. If we substitute $ct$ for $t$ in (8), the corresponding operator changes into

$$\frac{E}{c} \to i\frac{\hbar}{c} \frac{\partial}{\partial t} \tag{9}$$

From (8) and (9), we then obtain a tangent vector space on an almost complex manifold, which bases are

$$\frac{\partial}{\partial z^t} = i\frac{\hbar}{c} \frac{\partial}{\partial t} \quad , \quad \frac{\partial}{\partial z^\alpha} = -i\hbar \frac{\partial}{\partial \alpha} \tag{10}$$

Comparing (8) with (10), we see that the $\partial/\partial z^t$ and $\partial/\partial z^\alpha$ associate with $E/c$ and $P_\alpha$, $\alpha = x, y, z$, respectively. As a dual relation the bases of its dual space, a cotangent space, are

$$dz^t = i\, cdt \quad , \quad dz^\alpha = -i\, d\alpha \tag{11}$$

There is no the constant $\hbar$ in (11) since position and time have not been quantized in the quantum mechanics.

There exists, in addition, a pair of conjugate dual spaces relating to the conjugate equation (2.2), and their bases are defined as follows

$$\frac{\partial}{\partial \bar{z}^t} = -\mathrm{i}\frac{\hbar}{c}\frac{\partial}{\partial t} \quad , \quad \frac{\partial}{\partial \bar{z}^\alpha} = \mathrm{i}\hbar\frac{\partial}{\partial \alpha} \tag{12}$$

and

$$d\bar{z}^t = -\mathrm{i}\,cdt \quad , \quad d\bar{z}^\alpha = \mathrm{i}\,d\alpha \tag{13}$$

From (10)-(13) we then have two inner products of the two dual bases

$$\frac{\partial}{\partial Z^t} dz^t = \frac{\partial}{\partial \bar{z}^t} d\bar{z}^t = -\hbar \quad , \quad \frac{\partial}{\partial z^\alpha} dz^\alpha = \frac{\partial}{\partial \bar{z}^\alpha} d\bar{z}^\alpha = -\hbar \tag{14}$$

and their models are

$$\frac{\partial}{\partial z^t}\frac{\partial}{\partial \bar{z}^t} = \left|\frac{\partial}{\partial z^t}\right|^2 \quad , \quad \frac{\partial}{\partial z^\alpha}\frac{\partial}{\partial \bar{z}^\alpha} = \left|\frac{\partial}{\partial z^\alpha}\right|^2 \tag{15}$$

$$dz^t d\bar{z}^t = \left|dz^t\right|^2 \quad , \quad dz^\alpha d\bar{z}^\alpha = \left|dz^\alpha\right|^2 \tag{16}$$

Furthermore, from (14)-(16), we have

$$\left|\frac{\partial}{\partial z^t}\right|^2 \left|dz^t\right|^2 = \left|\frac{\partial}{\partial z^\alpha}\right|^2 \left|dz^\alpha\right|^2 = \hbar^2 \tag{17}$$

where the orthonormal condition (7) is used.

We now see that there are two almost complex manifolds, on each of which there are a tangent vector space and a cotangent vector space, and they relate to a momentum-energy and a space-time, respectively. It is well-known that in a momentum representation on a real manifold the momentum-energy is a cotangent vector space, at present in the coordinate representation on the almost complex manifold, however, we realize that the momentum-energy is of a tangent vector space, furthermore, between it and the space-time there is a dual relation which has never been seen on the real manifold. We think that this phenomenon results from the quantum mechanics itself, and it should show us some special properties for the wave-particle duality.

## 3. Results and discussion

### 3.1. A spatial meaning of the minimal uncertainty relations

In the uncertainty relations each of the variables $\Delta t, \Delta x, \Delta y, \Delta z$ and the $\Delta E, \Delta P_x, \Delta P_y, \Delta P_z$ is an observable physical variable of real number. By differential geometry, there is no metric significance for a space until a notion of metric is introduced[21]. In a 4-dimensional space on a real manifold a quadrate of an interval of an observable variable in one dimension equals to a product of a quadrate of the basis of the dimension and its metric[22]. For this reason, by (8), (9), (15) and (16) a relation among a mode of a basis, a metric and an observable-variable interval is expressed as

$$-g^{tt}\left|\frac{\partial}{\partial z^t}\right|^2 = (\frac{\Delta E}{c})^2 \quad , \quad g^{\alpha\alpha}\left|\frac{\partial}{\partial z^\alpha}\right|^2 = (\Delta P_\alpha)^2 \tag{18}$$

where the metrics $g^{tt} = -1, g^{\alpha\alpha} = 1$. Similarly,

$$-g_{tt}\left|dz^t\right|^2 = (c\Delta t)^2 \quad , \quad g_{\alpha\alpha}\left|dz^\alpha\right|^2 = (\Delta\alpha)^2 \tag{19}$$

where $g_{tt} = -1, g_{\alpha\alpha} = 1$, $\alpha = x, y, z$. We now explain further the meaning of $\Delta E$ in (18). As in the relativity theory, the $\Delta E$ is, first of all, an energy interval. Since the origin of the tangent vector space $\{\partial/\partial z^t, \partial/\partial z^\alpha, \alpha = x, y, z\}$ can be built up at any points where a particle exists, hence the $\Delta E$ associates with a random value of energy for the particle. Its magnitude is, however, independent of the magnitude of the particle-state energy, since the derivation of (18) is independent of an actual state. In the quantum mechanics sense, such that it should be regarded as a root-mean-square value about an average energy of a state, that is the $\Delta E$ is then statistically considered as a fluctuation of a random energy deviating from a mean one in a state. Similarly, the $\Delta P_\alpha$ and the $\Delta \alpha$ are of the same nature.

From (17), (18) and (19) we immediately get the minimal uncertainty relations

$$(-g^{tt})\left|\frac{\partial}{\partial z^t}\right|^2 (-g_{tt})\left|dz^t\right|^2 = (\Delta E)^2 \cdot (\Delta t)^2 = \hbar^2 \tag{20}$$

$$g^{\alpha\alpha}\left|\frac{\partial}{\partial z^\alpha}\right|^2 g_{\alpha\alpha}\left|dz^\alpha\right|^2 = (\Delta P_\alpha)^2 \cdot (\Delta\alpha)^2 = \hbar^2 \tag{21}$$

The correspondence of the (20) and (21) with the minimal uncertainty relations shows again us that the $\Delta E, \Delta P_\alpha$ and the $\Delta\alpha$ are each of them the root-mean-square values. The (20) and (21) reveal to us the restricted relations appearing in the minimal uncertainty relations are actually the restricted ones between these two 4-dimensional spaces on the almost complex manifold. We note that if there were a real part in each equation of the (10)-(13), there needed a normalizing factor, a real number 1/2, in each of the (10)-(13), respectively[21]. In fact, however, there exist merely the imaginary parts in these two spaces, and if we kept the factor, the real number 1/2, the right side of (17) would equal to $\hbar^2/4$ such that the magnitude of the minimal uncertainty relation would become $\hbar^2/4$, instead of $\hbar^2$. Of course, the results $\hbar^2$ and $\hbar^2/4$ are actually accepted for us.

**3.2. Local conservation and local equilibrium**

Mentioning the special relativity, Feynman explains that the 4-dimensional momentum-energy relation is actually a conservation law for momenta and energy[23]. Of course, this law has only a local meaning. On the other hand, the famous Chinese mathematician S. S. Chen points out that with the help of a fundamental tensor, we may identify a tangent space with a cotangent space, and hence a contravariant vector and a covariant vector can be viewed as different expressions of the same vector[21]. Combining these two opinions, we may then say that there should locally exist a space-time conservation with a momentum-energy one since their duality, that is, both of an infinitesimal increment of time and an infinitesimal increment of space obey commonly a conservation equation, a formula for space-time intervals[23], in a microscopic sense.

Like in the special relativity, if the origin of the coordinate system $(t, x, y, z)$ built up at a particle itself, and then its momenta and positions will vanish with their mean values of zero in such a reference-coordinate system, thus (21) will not exist and (20) will simply become

$$(-g^{tt})\left|\frac{\partial}{\partial z^t}\right|^2 (-g_{tt})\left|dz^t\right|^2 = (\frac{E_0}{c})^2 (c \cdot d\tau)^2 = \hbar^2 \tag{22}$$

where $(E_0)^2 = (\Delta E_0)^2 = \overline{(E_0 - \overline{E})^2}$, $E_0$ is the particle rest energy, $\overline{E} = 0$ is a mean energy and $d\tau$ is the proper time of the quantum state for the particle. Substituting $m_0 c^2$ for $E_0$ in (22), we then have

$$d\tau = \frac{\hbar}{E_0} = \frac{\hbar}{m_0 c^2} \tag{23}$$

where $m_0$ is the rest mass of the particle. While reviewing Bohr's reasoning in the Bohr-Einstein debate on the photon box experiment, by the Bohr's reasoning the authors of the reference [24] proved a minimal uncertainty relation between the proper time and the rest mass, which essentially satisfies the (23), where a constant is $h$ instead of $\hbar$. The significance of (23) is that each proper time for any quantum states of a particle with a rest mass $m_0$ is of the same constant, that is, the living time in each state for a particle keeps constant. If a quantum system consists of these identical particles, exchanging nothing with outside, the system will certainly be in a thermodynamic equilibrium state due to the Ergodic hypothesis[20]. In fact, this concept has been already applied in the classical quantum statistics without any verification[20]: A phase volume occupied by a quantum state in each degree of freedom in a statistical phase space consisting of position coordinates and momentum coordinates equals to the constant $\hbar$. The theory is referred to, perhaps, the following two causes. The first, the fluctuations will become the smallest when a system is in the equilibrium state ($\Delta x, \Delta P_x, \cdots$, each is a root-mean-square value). The second, considering a quantum effect, the restrictions of the uncertainty relations to these fluctuations should be true. The (23) is locally established. Even if the whole system is in a stable nonequilibrium state, some local equilibrium states will exist because of (23), which accords with the nonequilibrium statistical mechanics theory, that is, there always exist local equilibrium states in a stable thermodynamic system with a nonequilibrium state[25].

Exploring the large scale structure of space-time, S. W. Hawking uses the local conservation of energy and momentum, and calls it one of two postulates on the nature of a matter field equation, which is common to both the special and general theory of relativity[26]. If the postulate is reasonable, we may say, at least, that the local conservation of space and time will also hold for the theory since the duality of these two 4-dimensional spaces under quantum mechanics.

**3.3. An invariability under the special relativity**

Let there be two reference coordinate systems $S(t,x,y,z)$ and $S'(t',x',y',z')$ on the real manifold, and they be consistent with the Lorentz transformation[22]

$$x' = \gamma(x - vt) \quad , \quad x = \gamma(x' + vt') \tag{24}$$

$$t' = \gamma(t - \frac{\beta}{c}x) \quad , \quad t = \gamma(t' + \frac{\beta}{c}x') \tag{25}$$

where $\gamma = (1 - \beta^2)^{-1/2}$, $\beta = v/c$, $v$ be a velocity of the $S'$ with respect to the $S$ along the $x(x')$ direction. By differential chain rule, we have

$$\frac{\partial}{\partial x} = \gamma(\frac{\partial}{\partial x'} - \frac{\beta}{c}\frac{\partial}{\partial t'}) \quad , \quad \frac{\partial}{\partial t} = \gamma(\frac{\partial}{\partial t'} - v\frac{\partial}{\partial x'}) \tag{26}$$

and

$$dx = \gamma(dx'+vdt') \quad , \quad dt = \gamma(dt'+\frac{\beta}{c}dx') \qquad (27)$$

It is easy to prove

$$\frac{\partial}{\partial x}dx = \frac{\partial}{\partial x'}dx' = 1 \quad , \quad \frac{\partial}{\partial t}dt = \frac{\partial}{\partial t'}dt' = 1 \qquad (28)$$

Obviously, all of the formulas before (24) refer to the system S. Similarly, by the (28) we can simply verify that in the system S' the minimal uncertainty relations will have the same forms as (20) and (21), which shows that the law is of invariability under the special relativity.

### 3.4. A supposition about something having dark property

The state $\Psi$ and the state $\Psi^*$ should be regarded as two different states with the same energy on the almost complex manifolds, since both of them refer to two complex spaces with their distinguishable bases, respectively. Under the reversal symmetry of time these states of a particle exist simultaneously such that we can measure the particle with a certainty probability density $|\Psi|^2$ in the real space. A crucial cause resulting in a phenomenon of something having dark property is, perhaps, a breakdown of time symmetry. Landau indicated that in quantum mechanics there is a physical non-equivalence of the two directions of time, and theoretically the law of increase of entropy might be its macroscopic expression[25]. In general, that a motion is reversible with respect to time is usually formulated under the name of the principle of microreversibility. It is currently assumed that any quantum system evolving in the absence of external fields satisfies the microreversibility principle[12].

In the presence of an external field the principle will be, however, not satisfied if the external field changes with the time reversal. If there does not exist a local equilibrium state anywhere and anytime, that is, the principle of microreversibility will not hold, such that the time arrow has a certain direction for this system, and we then call this case a breakdown of time symmetry. When the case is true, the conjugate equation (2.2) referring to the reverse of time will not exist, and therefore there will not be the conjugate state $\Psi^*$. It is well-known that a mode $|\Psi|^2$ of a wave function $\Psi$ represents a probability density of existing in the space $(t, x, y, z)$ for a particle. In a mathematics sense, there will not be a mode of a complex variable unless there exists the conjugate one. Thus, there is no the $\Psi^*$, there is no the $|\Psi|^2$ (the wave function $\Psi$ is assumed to be complex valued here) such that there is no the probability density for a particle and the (3.2), (4),(5) and (6)will not exist, we will then not be able to measure any particle's dynamical variables appearing usually as their mean values, and the particle itself, even though it exists exactly. Furthermore, on the almost complex manifolds there will not be the conjugate dual spaces corresponding to (2.2), which bases are denoted by (12) and (13), then there will not be the modes such as the $\left|\partial/\partial z^t\right|^2$, $\left|\partial/\partial z^\alpha\right|^2$, $\left|dz^t\right|^2$ and the $\left|dz^\alpha\right|^2$, such that we will not be able to measure the momentum, energy, position and the time relating to (18) and (19). Finally, we will be faced with a dark matter, a dark momentum-energy and a dark space-time, the later two dark spaces will dually exist. Our supposition needs to be further examined. Of course, this is an interesting topic.

## 4. Conclusion

Under quantum mechanics a momentum-energy and a space-time are a pair of dual vector spaces on an almost complex manifold in the position representation. The minimal uncertainty relations result from the restricted relations between the bases for these two 4-dimensional spaces, and the law will be invariable in the special relativity. In a closed system consisting of identical particles the relations associate with a thermodynamic equilibrium state. In a stable nonequilibrium state there are locally a momentum-energy conservation and a space-time conservation, which relate with local equilibrium states, in a microscopic sense. A supposition about existence of something having dark property is proposed: A breakdown of time symmetry results in that the principle of microreversibility will not hold, such that there will be no a conjugate state $\Psi^*$ for a state $\Psi$ of a particle, thus there will be no a probability density $\left|\Psi\right|^2$ of the particle in the state $\Psi$, and we will not be able to measure any dynamical variables of the particle and the particle itself even though it does exist. Our supposition needs to be further explored and examined.

## References

[1]  G. Rigolin 2000 Found. Phys. Lett. **15** 293-298

[2]  Paul S. Wesson 2004 Gen. Rel. Grav. **36** 451-457

[3]  Achin Kempf, Gianpiero Mangano and Robert B. Mann 1995 Phys. Rev. **D52** 1108-1118

[4]  M. T. Jaeckel and S. Reynaudn 1994 Phys. Lett. **A185** 143

[5]  Luis J. Garay 1995 Int. Mod. Phys. **A10** 145-166

[6]  Lay Nam Chang et. al. 2002 Phys. Rev. **D65** 125028

[7]  T. Yoneya 1989 Mod. Phys. Lett. **A4** 1587

[8]  M. Li and T. Yoneya 1997 Phys. Rev. Lett. **78** 1219

[9]  T. Yoneya 1997 arXiv: hep-th/9703078

[10]  Ichiro Oda 1998 Mod. Phys. Lett. **A13** 203-210

[11]  T. Yoneya 2000 Prog. Theor. Phys. **103** 1081-1125

[12]  A. Messiah 1999 Quantum Mechanics, Two Volumes Bound as One (New York: Dover Pub. Inc.) p258, 674-675

[13]  W.Pauli 1980 General Principles of Quantum Mechanics (New York: Springer-Verlag) p3, 136, 319

[14]  Muga J. G. and Leavens C. R. 2000 Phys. Rep. **338** 353

[15]  Bohr N. 1983 Quantum Theory and Measurement eds. J. A. Wheeler and W. H. Zurek (Princeton N. J.: Princeton Uni. Press.) p9-49

[16]  Romeo Brunetti and Klaus Fredenhag 2002 Rev. Math. Phys. **14** 897-906

[17]  L. Burakovsky, L. P. Horwitz and W. C. Schieve 1997 Found. Phys. Lett. **10** 503-516

[18]  Z. Y. Wang, B. Chen and C. D. Xiong 2003 J. Phys. A: Math. Gen. **36** 5136-5147

[19]  Hitoshi Kitada, Komaba and Meguro 2000 arXiv:quant-ph/0007028

[20]  R. K. Pathria 1997 Statistical Mechanics, 2-end. Edit. (Oxford: Butterwoth Heinmann) p33, 41, 39

[21]  S. S. Chen, W. H. Chen and K. S. Lam 2000 Lectures on Differential Geometry (Sinhapore: World Scientific Pub. Co. Pte. Ltd.) p227-233, 238,130,136

[22]  W. Pauli 1981 Theory of Relativity (New York: Dover Pub.) p27-28,10

[23]  R.Feynman, Robert B. Leighton and Matthew L. Sands 2004 Lectures on Physics, Vol. 1 (Beijing: World Publishing Corp.) p17-7, 17-3

[24]  Shoju Kudaka 1999 J. Math. Phys. **40** 1237-1245

[25]  L. D. Landau and E. M. Lifshitz 1999 Statistical Physics, Part 1, 3-rd. edit.

(Oxford: Butterwoth-Heinemann) p27, 32-33

[26] S. W. Hawking and G. F. R. Ellis 1999 The Large Scale Structure of Space-time (Cambridge: Cambridge Uni. Press.) p59-61